\documentclass[12pt]{iopart}

\usepackage{envmath}
\usepackage{graphicx}

\usepackage[latin1]{inputenc}
\usepackage{amsfonts}
\usepackage{amssymb}

\usepackage{graphicx}
\usepackage{subfigure}

\usepackage{url}
\usepackage{mathrsfs}

\usepackage{xcolor}

\def \M{{\cal M}}

\def \a{\alpha}
\def \Fs{{\cal F}}

\def \A{{\cal A}}
\def \B{{\cal B}}
\def \J{{\cal J}}

\def \x{{\bf x}}
\def \y{{\bf y}}
\def \h{{\bf h}}

\def \Ac{{\bf A}}
\def \Bc{{\bf B}}
\def \S{{\bf S}}

\def \Y{{\bf Y}}
\def \N{{\bf N}}
\def \U{{\bf U}}
\def \hf {\frac{1}{2}}
\def \qtr {\frac{1}{4}}
\def \tx {\tilde {x}}
\def \ty {\tilde {y}}

\def \no {\nonumber}
\def\lsim{\mathrel{\rlap{\lower4pt\hbox{\hskip1pt$\sim$}}
    \raise1pt\hbox{$<$}}}                % less than or approx. symbol
\def\gsim{\mathrel{\rlap{\lower4pt\hbox{\hskip1pt$\sim$}}
    \raise1pt\hbox{$>$}}}                % greater than or approx. symbol

\begin{document}

\title{Marginalizing the likelihood function for modeled gravitational wave searches} 

\author{Sanjeev Dhurandhar$^{1,2}$, Badri Krishnan$^{2,4}$ and Joshua L.~Willis$^{3,2}$} 

\address{$^1$ Inter-University Center for Astronomy and Astrophysics,
  Post Bag 4, Ganeshkhind, Pune 411007, India}

\address{$^2$ Max Planck Institute for Gravitational Physics (Albert
  Einstein Institute), Callinstr. 38, 30167 Hannover, Germany }

\address{$^3$ Abilene Christian University, Abilene, TX 79699, USA}

\address{$^4$ Leibniz Universit\"at Hannover, D-30167 Hannover, Germany} 

\ead{badri.krishnan@aei.mpg.de}

\begin{abstract}

  Matched filtering is a commonly used technique in gravitational wave
  searches for signals from compact binary systems and from rapidly
  rotating neutron stars.  A common issue in these searches is dealing
  with four extrinsic parameters which do not affect the phase
  evolution of the system: the overall amplitude, initial phase, and
  two angles determining the overall orientation of the system.  The
  $\Fs$-statistic maximizes the likelihood function analytically over
  these parameters, while the $\B$-statistic marginalizes over them.
  The $\B$-statistic, while potentially more powerful and capable of
  incorporating astrophysical priors, is not as widely used because of
  the computational difficulty of performing the marginalization.  In
  this paper we address this difficulty and show how the
  marginalization can be done analytically by combining the four
  parameters into a set of complex amplitudes.  The results of this
  paper are applicable to both transient non-precessing binary
  coalescence events, and to long lived signals from rapidly rotating
  neutron stars.

\end{abstract}


\section{Introduction}
\label{sec:intro}

Gravitational wave (GW) searches rely on matched filtering in
situations where the expected gravitational wave signal is well known.
This is true for both the transient broad-band signals from compact
binary coalescence (CBC) events, and for the long lived but
narrow-band continuous wave (CW) signals expected from rapidly
rotating neutron stars.  For both these cases, as we shall explain in
more detail later, there are four parameters which do not affect the
phase evolution of the system and are independent of the gravitational
wave detector(s): the overall amplitude of the signal $A$, the initial
phase of the signal $\varphi_0$, the polarization angle $\psi$, and
the inclination angle of the axis of symmetry of the system $\iota$
\footnote{The analysis in this paper is restricted to non-precessing
  systems for which the inclination angle is constant in time.}
relative to the line of sight between the system and the
detector. There will be other parameters, such as the sky-position,
intrinsic parameters such as the masses and spins in a binary system,
and the signal frequency and spindown in the case of an isolated
neutron star.  These parameters may or may-not affect the phase
evolution of the signal as seen by a detector, but in any case
$(A,\varphi_0, \psi,\iota)$ will not do so.  The phase evolution on
the other hand, is determined by parameters of arguably greater
physical interest such as the masses and spins of the components of a
binary system, and in fact $(A,\varphi_0, \psi,\iota)$ are sometimes
referred to as ``nuisance parameters''.  We would prefer to not have
to explicitly search over $(A,\varphi_0, \psi,\iota)$ and to deal with
them analytically as far as possible.

There are two known methods for dealing with such ``nuisance''
parameters both of which are based on the likelihood function
$\Lambda(A,\varphi_0,\psi,\iota|\mathbf{x})$. Here $\mathbf{x}$ is a
given data series from a network of GW detectors, and $\Lambda$ is
defined as the ratio
\begin{equation}
  \Lambda(A,\varphi_0,\psi,\iota|\mathbf{x}) := \frac{p(\mathbf{x}|A,\varphi_0,\psi,\iota)}{p(\mathbf{x}|\mathbf{noise})}\,.
\end{equation}
The numerator is the probability density for the data $\mathbf{x}$ for
a specific value of the nuisance parameters, while the denominator is
the same probability density function in the absence of a signal.  We
have suppressed the dependence of $\Lambda$ on other physical source
parameters. The first and more commonly used method, is to maximize
$\Lambda$ over the nuisance parameters; this is the standard
prescription in the Frequentist framework.  This is ultimately
motivated by the Neyman-Pearson lemma (see e.g. \cite{kendall2A})
which is applicable when we wish to distinguish between two simple
hypotheses; here the two hypotheses corresponds to the cases when i)
there are no signals present (the null hypothesis), and ii) when we
have a signal with fixed values of the nuisance parameters.  Consider
the problem of finding a region $\mathcal{R}$ in the space of data
vectors $\mathbf{x}$ such that the probability of detections is
maximized at a fixed false alarm probability.  The Neyman-Pearson
lemma tells us that level surfaces of the likelihood function yield
the most powerful test, i.e. the best choice of the region
$\mathcal{R}$ with the largest detection probability for a fixed false
alarm probability. For composite hypotheses where all allowed values
of the amplitude are included, there is no known analog of the
Neyman-Pearson lemma.  However, the Neyman-Pearson lemma still
motivates us to look at the likelihood function, and to find the
values of the nuisance parameters which maximizes it.

In practice for GW searches, one combines the nuisance parameters into
quantities (the so-called amplitude parameters) which appear linearly
in the GW signal model.  This linearity enables one to maximize
$\Lambda$ analytically over the amplitude parameters; in the
gravitational wave data analysis context, this was first shown in 1998
\cite{Jaranowski:1998qm} for CW signals and soon generalized to the
coherent multi-detector case \cite{Cutler:2005hc}, and for CBC
searches \cite{Pai:2000zt, Harry:2010fr, Keppel:2013uma}.  The
analytic maximization allows one to focus computational resources on
numerically maximizing $\Lambda$ over the other signal parameters
determining the phase evolution of the signal.

The second method is to marginalize $\Lambda$ over the nuisance
parameters, i.e. to compute the integral
\begin{equation}
  \int \Lambda(A,\varphi_0,\psi,\iota|\mathbf{x}) p(A)\, dA\, d\varphi_0\, d\psi\, d\cos\iota\,.
\end{equation}
Here $p(A)$ is an astrophysical prior on the amplitude.  For CBC
searches this depends primarily on the assumed spatial distribution of
sources, while for CW signals this depends also on properties of
neutron stars such as their deformations away from axisymmetry.  The
volume element corresponds to an isotropic distribution of the angles
$(\varphi_0,\psi,\iota)$ which is realistic when we have no other
prior information about the system.  The Bayesian approach has some
benefits such as allowing one to incorporate physical priors, and is
potentially more powerful than the $\Fs$-statistic.  Moreover, in the
gravitational wave literature it was shown in \cite{Searle:2008jv}
that if the prior used in the marginalization is in fact the true
distribution found in nature, then indeed, the marginalized likelihood
function is the optimal detection statistic in the Neyman-Pearson
sense (see also Sec. 21.28 of \cite{kendall2A}).

However, the above integral is not easy to compute analytically which
makes this method less useful in practice.  This was first proposed as
a detection statistic for CW searches in \cite{Prix:2009tq}, which
defined the so called $\B$-statistic (see also \cite{Searle:2008jv}).
This work however did not address the critical question of
computational cost in evaluating the integral.  This was partially
addressed in \cite{Whelan:2013xka} where a different parametrization
of the amplitude parameters was constructed based on left- and
right-circular polarizations.  However useful approximations to the
$\B$-statistic integral were found only in the linearly polarized case
(where $\cos\iota=0$).  These coordinates were used for CBC searches
in \cite{Haris:2016jap} which obtained analytic approximations to the
integral above, and presented an initial numerical comparison of the
Bayesian and Frequentist approaches. Here we propose a new set of
complex coordinates (closely related to
\cite{Whelan:2013xka,Haris:2016jap}) for the space of amplitude
parameters motivated by properties of the rotation group, which works
for both CW and CBC waveforms.  These coordinates allow for a more
detailed understanding the singularities of the $\B$-statistic and
also pave the way for a generalization to include higher modes of the
waveform and e.g. precession effects.  The same approach also allows
for a better understanding of the detector beam pattern functions.

A central message of this paper is the following.
$(\varphi_0,\psi,\iota)$ are best thought of as points on a 3-sphere
and we can thus identify the 3-sphere with the group of rotations in
three dimensional Euclidean space.  Given the appropriate tensorial
structure of the fields of interest, i.e. fields which transform under
rotations as a representation of weight $\ell$, we should use the
corresponding matrix elements of the rotation group to expand
functions of $(\varphi_0,\psi,\iota)$.  For gravitational waves, we
are dealing with the $\ell=2$ case and this determines the appropriate
basis functions that we should be using to expand functions of
$(\varphi_0,\psi,\iota)$.  While this may be interesting from a
theoretical viewpoint, there is \emph{apriori} no guarantee that this
should help simplify the calculation of the $\B$-statistic, but in
fact, as we shall show, it does do so.

In the following sections, we start in Sec.~\ref{sec:preliminaries}
with a review of the waveform model, the response of a GW detector and
the $\Fs$- and $\B$-statistics.  The new complex amplitudes are
defined in Sec.\ref{sec:newamplitudes}. Sec.~\ref{sec:likelihood}
demonstrates how the likelihood function is expressed in terms of the
new amplitudes.  Sec.~\ref{sec:evaluatingB} shows how $\Lambda$ can be
analytically marginalized over the complex amplitudes along with some
example cases and we finally conclude in Sec.~\ref{sec:discussion}.

\section{Preliminaries and notation}
\label{sec:preliminaries}

Consider a plane gravitational wave (GW) $h_{ab}$ and an associated
right-handed orthonormal wave-frame
$(\mathbf{X},\mathbf{Y},\mathbf{Z})$ such that the wave is traveling
along the $\mathbf{Z}$ direction. In standard linearized general
relativity, gravitational waves have two polarizations states defined
by the symmetric-tracefree tensors
\begin{equation}
  (\mathbf{e}_+)_{ab} = X_aX_b - Y_aY_b\,,\quad (\mathbf{e}_\times)_{ab} = X_aY_b + Y_aX_b\,.
\end{equation}
Then we can write $h_{ab}$ as
\begin{equation}
  h_{ab} = h_+(\mathbf{e}_+)_{ab} + h_\times (\mathbf{e}_\times)_{ab}\,.
\end{equation}
In this paper we shall use instead the complex null-vector
\begin{equation}
  \mathbf{m} = \frac{1}{\sqrt{2}}(\mathbf{X} + i\mathbf{Y})\,.
\end{equation}
It is easy to see that $\mathbf{m}\cdot \mathbf{m} = 0$ and
$\mathbf{m}\cdot \mathbf{m}^\star = 1$.  In general the gravitational
wave $h_{ab}$ can then be written as
\begin{equation}
  \label{eq:hab}
  h_{ab} = \mathfrak{h}^\star m_am_b + \mathfrak{h} m^\star_am^\star_b\,,
\end{equation}
where the complex scalar $\mathfrak{h}$ contains the two
polarizations $h_{+,\times}$:
\begin{equation}
  \mathfrak{h} = h_+ + ih_\times\,.
\end{equation}
Using $m_a$ instead of $X_a$ and $Y_a$ is equivalent to using left and
right circular polarizations instead of linear polarization states.
Note that if we perform a counter-clockwise rotation in the
$(\mathbf{X},\mathbf{Y})$ plane by an angle $\psi$:
\begin{eqnarray}
  \mathbf{X} &\rightarrow& \cos\psi \mathbf{X} + \sin\psi\mathbf{Y}\,,\\
  \mathbf{Y} &\rightarrow& -\sin\psi \mathbf{X} + \cos\psi\mathbf{Y}\,,
\end{eqnarray}
then $\mathbf{m}\rightarrow e^{-i\psi}\mathbf{m}$, and thus
$\mathfrak{h} \rightarrow e^{-2i\psi}\mathfrak{h}$; $m_a$ is, by
definition, assigned a spin weight +1, and $\mathfrak{h}=m^am^bh_{ab}$
is said to have spin-weight +2.  For an elliptically polarized
wave, it is always possible to choose the wave frame aligned with the
principal directions of the polarization ellipse.  In this case $h_+$
and $h_\times$ have a $\pi/2$ offset in phase:
\begin{eqnarray}
  \label{eq:hpc}
  h_+(t) = A_+(t)\cos\Phi(t) \,,\qquad h_\times(t) = A_\times(t)\sin\Phi(t)\,.
\end{eqnarray}
Here the amplitudes $A_{+,\times}$ are slowly varying functions of
time while the phase $\Phi(t)$ is rapidly varying.  We shall assume
that the wave-frame is aligned with the principal polarization
directions in this way.

Consider now an interferometric GW detector on Earth.  We shall work
in the long-wavelength approximation (appropriate for current ground
based detectors) where the GW wavelength $\lambda$ is much larger than
the interferometer arm-length.  In this approximation, the strain
$h(t)$ measured by the detector will be given as
\begin{equation}
  \label{eq:hD}
  h(t) = h_{ab}D^{ab} 
\end{equation}
where $D_{ab}$ is the detector tensor
\begin{equation}
  D_{ab} = \frac{1}{2}(u_au_b - v_av_b)\,.
\end{equation}
(we refer the reader to e.g. \cite{1987MNRAS.224..131S} for further
details, and to e.g. \cite{EstabrookWahlquist1975, Finn:2008np} when
it is necessary to go beyond the long-wavelength approximation).  Here
$\mathbf{u},\mathbf{v}$ are unit-vectors along the detector arms, and
we follow the convention that $(\mathbf{u},\mathbf{v},\mathbf{z})$
forms a right handed coordinate system with $\mathbf{z}$ pointing away
from Earth's center.  Let us also take the angle between $\mathbf{u}$
and $\mathbf{v}$ to be $2\zeta$.  Take a frame
$(\mathbf{x},\mathbf{y})$ such that $\mathbf{x}$ bisects the arms of
the detector.  In this case
$D_{ab} = \frac{1}{2}\sin 2\zeta (x_ay_b + y_ax_b)$.  Unless mentioned
otherwise, we henceforth take the arms to be perpendicular and
keep in mind that an overall factor of $\sin2\zeta$ suffices to
account for it if necessary.

Introduce spherical polar coordinates $(\theta,\phi)$ associated with
the detector frame $(\mathbf{x},\mathbf{y},\mathbf{z})$ in the usual
way, and the corresponding unit vectors
$(\mathbf{e}_\theta,\mathbf{e}_\phi)$.  The direction of propagation
of the GW, i.e. $\mathbf{Z}$ is pointed towards the detector.  We can
take $(\mathbf{e}_\theta, -\mathbf{e}_\phi,\mathbf{Z})$ as a reference
wave frame and
$\mathbf{m}^{(0)} = (\mathbf{e}_\theta- i\mathbf{e}_\phi)/\sqrt{2}$ as
a reference null spin vector.  The wave-frame $(X,Y)$ will be related
to $(-\mathbf{e}_\theta,\mathbf{e}_\varphi)$ by a counter-clockwise
rotation (as determined by $\mathbf{Z}$ being inward pointing).  This
implies that we can rotate $\mathbf{m}^{(0)}$ and align it with
$\mathbf{m}$: $\mathbf{m}= e^{-i\psi}\mathbf{m}^{(0)}$.  With these
conventions, we can rewrite $h_{ab}$ from Eq.~(\ref{eq:hab}) as
\begin{equation}
  \label{eq:hab-m0}
  h_{ab} = \mathfrak{h}^\star e^{-2i\psi} m^{(0)}_am^{(0)}_b + \mathfrak{h} e^{2i\psi} m^{(0)\star}_am^{(0)\star}_b \,.
\end{equation}
The detector response can be written in terms of a complex beam
pattern function $\mathfrak{F}(\theta,\phi,\psi)$ as
\begin{equation}
  \label{eq:hf}
  h(t) = \frac{1}{2}\mathfrak{h} \mathfrak{F}^\star(\theta,\phi,\psi) + \frac{1}{2}\mathfrak{h}^\star \mathfrak{F}(\theta,\phi,\psi)
\end{equation}
with $\mathfrak{F} = F_+ + iF_\times$.  The factor of 1/2 has been
chosen so that the above is equivalent to
$h = F_+h_+ + F_\times h_\times$.  From Eqs.~(\ref{eq:hab-m0}) and
(\ref{eq:hD}) we get
\begin{eqnarray}
  \label{eq:complexF}
  \mathfrak{F}(\theta,\phi,\psi) &=& 2D^{ab}m_am_b = 2e^{-2i\psi}D^{ab}m^{(0)}_am^{(0)}_b \\
                                 &=& \left( \frac{1+\cos^2\theta}{2} \sin 2\phi - i\cos\theta \cos2\phi\right) e^{-2i\psi}\,.
\end{eqnarray}
We often write Eq.~(\ref{eq:complexF}) as
\begin{equation}
  \label{eq:detresponse}
  \mathfrak{F}(\theta,\phi,\psi) = \mathfrak{f}(\theta,\phi)e^{-2i\psi}\,.
\end{equation}
This separates the polarization angle $\psi$ from the sky-position.
As in \cite{Jaranowski:1998qm}, it
is useful to define the $\psi$ independent beam pattern functions
$a(\theta,\phi)$ and $b(\theta,\phi)$ as
\begin{equation}
  \mathfrak{f}(\theta,\phi) = a(\theta,\phi) + ib(\theta,\phi)\,.
\end{equation}
This leads to
\begin{equation}
  \label{eq:ab}
  F_+ = a\cos 2\psi + b \sin 2\psi\,,\quad F_\times = b\cos 2\psi - a \sin 2\psi\,.
\end{equation}
When dealing with multiple detectors, one refers $(\theta,\phi,\psi)$
to a common coordinate system such a geocentric or a solar system
barycenter based system. In this case we can still separate the
sky-positions from the polarization angle. 

The time parameter $t$ appearing in Eq.~(\ref{eq:hf}) is the arrival
time of wavefronts at the detector.  It is related to the time of
arrival at the origin of coordinates $\tau$ according to
$t = \tau - \mathbf{r}\cdot\mathbf{n}/c$ where $\mathbf{r}$ is the
position of the detector, and $\mathbf{n}$ the direction to the source
and $c$ the speed of light. For transient signals it is common to use
a geocentric system while for CW sources a solar-system barycentric
system is more natural.  For long duration signals this leads to the periodic
Doppler shifting of the signal frequency as Earth rotates around its
axis and orbits the Sun, and furthermore, it is often necessary to
take into account the relativistic Einstein and Shapiro time delays
\cite{Jaranowski:1998qm}.

\subsection{The relation between the beam pattern functions and representations of the rotation group}
\label{subsec:beam-group}

It is interesting to note that, as first done in
\cite{1988MNRAS.234..663D}, $\mathfrak{F}$ can be written in the
following way:
\begin{equation}
  \label{eq:complex-beam}
  \mathfrak{F} = ie^{-2i\phi}\frac{(1-\cos\theta)^2}{4}e^{-2i\psi} - ie^{2i\phi}\frac{(1+\cos\theta)^2}{4}e^{-2i\psi} \,.
\end{equation}
As shown in \cite{1988MNRAS.234..663D} the two terms appearing here
are essentially matrix elements of the rotation group, which in turn
are related to the spin-weighted spherical harmonics
\cite{Goldberg:1966uu,Newman:1961qr} (see \ref{sec:appendix}).  It is
in fact generally true that the detector response function can be
expanded in terms of the matrix elements of the rotation group. To see
this we briefly review some results from the representation theory of
the rotation group in 3-dimensional Euclidean space.  We later apply
the same approach to the emitted GW signal emitted by a CBC or CW
source.  We shall mostly follow \cite{GMS}.

Let $\mathcal{G}$ denote the group of rotations in $\mathbb{R}^3$ and
let $g\in \mathcal{G}$ be a particular rotation. Let the matrix
elements of $g$ be $g_{ij}$ (satisfying
$\mathbf{g}^\intercal\mathbf{g} = 1$) which transforms a vector $x_i$
as $x^\prime_i = \sum_{j=1}^3g_{ij}x_j$.  It will be convenient to use
the Euler angles to represent a general rotation.  We use the 'zxz'
convention where a general rotation is composed of three rotations: i)
a rotation around the $z$-axis by an angle $\varphi$, ii) a rotation
by an angle $\theta$ around the $x$-axis, and finally, iii) a rotation
by an angle $\psi$ around the $z$-axis.  Thus, we write
$g(\varphi,\theta,\psi)$ for a general rotation.  The inverse of
$g(\varphi,\theta,\psi)$ is seen to be
$g(\pi-\psi,\theta,\pi-\varphi)$.  We take all transformations
to be active.

Consider now an irreducible representation of $\mathcal{G}$ of weight
$\ell$.  Thus, we have a $2\ell+1$ dimensional complex vector space,
and each rotation $g$ is represented by a $2\ell+1$ dimensional
unitary matrix $T^\ell_{mn}(g)$, with $-\ell\leq m,n \leq \ell$.  We
shall often drop the superscript $\ell$ when it is obvious what the
weight is, and we shall also write $T^\ell_{mn}(\varphi,\theta,\psi)$,
i.e. as functions on $\mathcal{G}$.  The product of two rotations is
written as a matrix multiplication
\begin{equation}
  \label{eq:product}
  T^\ell_{mn}(g_1g_2) = \sum_{k=-\ell}^\ell T^\ell_{mk}(g_1)T^\ell_{kn}(g_2)\,.
\end{equation}
This equation can be viewed in the following way: under the
transformation $g\rightarrow g g_1$, the rotation group acts on the
matrix elements themselves.  Thus, each row of the matrix $T^\ell_{mn}$
(i.e. for fixed $m$) transforms as an irreducible representation of
the rotation group.  The general expression for the matrix elements
are summarized in \ref{sec:appendix}.

An important fact we shall use is the following: The functions
$T^\ell_{mn}(\varphi,\theta,\psi)$ form a complete orthogonal basis in
the Hilbert space of square integrable functions
$f(\varphi,\theta,\psi)$ on $\mathcal{G}$.  This fact can be used to
expand a general tensor on the 2-sphere $S^2$ (such as a gravitational
wave arriving from different sky-directions and with different
polarization angles) by the following construction.  Start with the
standard triad at the north-pole i.e. the basis vectors along the
$x,y,z$ directions.  Let us denote this triad as $\mathbf{e}_i$
($i=1,2,3$) and consider a gravitational wave arriving from a
sky-position P with coordinates $(\theta,\phi)$, and with a
polarization angle $\psi$. The rotation $g(\pi,\theta, \phi + \pi/2)$
takes the north pole to the point $(\theta,\phi)$ and the triad at the
north-pole is transformed to
$(-\mathbf{e}_\phi,\mathbf{e}_\theta,\mathbf{e}_r)$.  The inverse
rotation $g(\pi/2 - \phi,\theta,0)$ takes the point P to the north pole
and takes the wave-frame with $\psi=0$ to the triad at the north pole.
To go to the wave-frame with a polarization angle $\psi$, the
transformation is $g(\pi-\psi, \theta,\phi + \pi/2)$.  This
construction yields a triad for every group element, i.e. for all
values of $(\varphi,\theta,\psi)$.  By definition then, each of the
basis vectors then transforms as a scalar on $\mathcal{G}$ under the
transformation $g\rightarrow gg_1$. The components of any tensor field
in this basis are then to be viewed as functions of
$(\varphi,\theta,\psi)$ and thus can be expanded in terms of
$T^\ell_{mn}$.  An important ingredient here is the spin-weight: any
quantity with spin weight $s$ can be expanded in terms of
$T^\ell_{sm}$.  A scalar corresponds to $s=0$, and in fact the
functions $T^\ell_{0m}$ are just the usual spherical harmonics
$Y_{\ell m}$.  This is just the familiar observation that any scalar
function can be expanded in terms of the spherical harmonics.  For
gravitational waves, the relevant tensorial structure is that of
symmetric trace-free tensors and, we shall see that the appropriate
basis functions are $T^\ell_{-2,m}$.

Let us now return to the beam pattern function and write it in terms
of the matrix elements which are listed in \ref{sec:appendix}. The
relevant matrix elements of this transformation are
\begin{equation}
  T^\ell_{mn} (\pi-\psi,\theta,\phi+\pi/2) = (-1)^{m+n}i^n e^{im\psi}P^\ell_{mn}(\cos\theta)e^{-in\phi}\,. 
\end{equation}
The complex antenna pattern of Eq.~(\ref{eq:complex-beam}) can be
written as
\begin{equation}
  \label{eq:antenna-T22} 
  \mathfrak{F}(\theta,\phi,\psi) = i T^2_{-2,-2}(\pi-\psi, \theta,\phi+ \pi/2) - iT^2_{-2,2}(\pi-\psi, \theta,\phi+ \pi/2)  \,.
\end{equation}
If we go back to the more general expression
Eq.~(\ref{eq:detresponse}) for the beam pattern function, we see that
$\mathfrak{F}$ is best viewed as a function on $\mathcal{G}$ or,
equivalently, the 3-sphere $S^3$.  In addition, $\mathfrak{F}$ is seen
to have spin weight -2 because of the factor of $e^{2i\psi}$.  This
implies that in general $\mathfrak{F}$ can be expanded in terms of
$T^\ell_{-2,m}$.  Typically, only the $\ell=2$ components are
considered in the literature.  There are some exceptions such as
\cite{Baskaran:2003bx} which considers the $\ell=3$
contributions.  

As an example of the utility of this formalism, consider a situation
when we use a coordinate system not centered on the detector.  For a
detector on Earth, it is useful to consider a geocentric coordinate
system. Thus, let $(\Theta_I,\Phi_I)$ be the position of an
interferometric detector, and let $\Psi_I$ be the orientation of its
arms with respect to the lines of latitude.  Let $g$ be the
transformation from the geocentric system to the wave frame, and let
$g_I$ be the transformation from the geocentric system to the detector
frame.  Then, the transformation $g*g_I^{-1}$ takes us from the
detector to the wave-frame. From Eq.~(\ref{eq:product}) we get for
example
\begin{eqnarray}
  T^2_{-2,-2}(gg_I^{-1}) &=& \sum_{s=-2}^2  T^2_{-2,s}(g)T^2_{s,-2}(g_I^{-1}) \\ 
  &=& \sum_{s=-2}^2  T^2_{-2,s}(\pi-\psi,\theta,\phi+\pi/2)T^2_{s,-2}(\pi/2 - \Phi_I,\Theta_I,\Psi_I)  \,.
\end{eqnarray}
Viewed as an expansion on the sky, this is an expansion in the five
dimensional space spanned by $T^2_{-2,s}$ for $-2\leq s \leq 2$ and the
coefficients of the expansion depend on the location and orientation of
the detector. Using the expressions in \ref{sec:appendix}, it is
straightforward to obtain analytic expressions for these
coefficients and thus for the beam pattern functions.

\subsection{The $\Fs$- and $\B$-statistics}
\label{subsec:amplparams}

We now summarize the conventional approach to matched filtering in GW
data analysis using the so-called $\Fs$ and $\B$ statistics.  Let us
consider $M$ detectors labeled by an index $I = 1,2\ldots M$.  As in
Eq.~(\ref{eq:hpc}) taking $\Phi(t) = 2\varphi_0 + 2\varphi(t)$, we
write $h_{+,\times}$ in terms of a slowly varying amplitude $A(t)$ and
a rapidly varying phase $\varphi(t)$:
\begin{eqnarray}
  h_+ = A_+(t)\cos(2\varphi_0 + 2\varphi(t))\,,\nonumber\\
  h_\times = A_\times(t)\sin(2\varphi_0 + 2\varphi(t))\,.  
\end{eqnarray}
Here we include a factor of 2 with the phase as is natural for GW
sources, and $\varphi_0$ is an initial phase which depends on a choice
of frame attached with the source.  For CBC systems, $\varphi(t)$ is
the orbital phase of the binary system while in the CW case this is
the rotational phase of the neutron star.  For the dominant $\ell=m=2$
mode for non-precessing systems, including both CW and CBC sources, we
write the amplitudes $A_{+,\times}$ as
\begin{equation}
  A_+ = A\eta(t)\frac{1+\cos^2\iota}{2}\,,\\
  A_\times = A\eta(t)\cos\iota\,.
\end{equation}
Here $\eta(t)$ is a slowly varying function of time, $\iota$ is the
angle between the line of sight to the system and the system's axis of
rotation; for non-precessing systems $\iota$ would not be time
dependent.

For CW sources, $A_{+,\times}(t)$ are in fact constants over time and
are written in terms of an overall amplitude $h_0$.  Thus, $A=h_0$ and
$\eta(t)=1$.  The amplitude $h_0$ depends inversely on the distance
and additionally it depends on physical properties of say the neutron
star crust, the fluid motion in the interior etc.  For the
non-precessing CBC case, we have the same dependence on the angle
$\iota$.  In addition, the amplitude will depend on the masses and
spins of the binary components and the distance $D$ to the binary.
Furthermore, $\eta$ is no longer constant, but instead will increase
as the separation between the binary components decreases (see
e.g. \cite{Peters:1963ux}). It is convenient, as in
\cite{Harry:2010fr}, to write this as $A=D_0/D$ where $D_0$ is a
fiducial distance.

We can write $h_+$ and $h_\times$ as
\begin{eqnarray}
  h_+ = A_+(t)\cos 2\varphi_0\cos 2\varphi(t) - A_+(t)\sin 2\varphi_0 \sin 2\varphi(t) \,,\nonumber\\
  h_\times = A_\times(t)\sin 2\varphi_0\cos 2\varphi(t) + A_+(t)\cos 2\varphi_0 \sin 2\varphi(t)\,.
\end{eqnarray}
Combining $h_{+,\times}$ and the beam pattern functions
$F_{+,\times}$, and separating out the polarization angle explicitly,
it is easy to show that the signal in detector $I$ can be written as
\begin{equation}
  h^I (t) = \sum_{\mu=1}^4 \A^\mu h_\mu^I (t) \,, 
\label{eq:HF_basis}
\end{equation}
where $\A^\mu, ~\mu = 1, 2, 3, 4$ are amplitudes depending on the
distance to the source, initial phase $\varphi_0$, polarization angle
$\psi$ and the inclination angle $\iota$ of the source and are given
by:
\begin{eqnarray}
  \label{eq:amp}
  \A^1 &=& A_+ \cos 2 \varphi_0 \cos 2 \psi - A_\times \sin 2 \varphi_0 \sin 2 \psi \,, \no \\
  \A^2 &=& A_+ \cos 2 \varphi_0 \sin 2 \psi + A_\times \sin 2 \varphi_0 \cos 2 \psi \,, \no \\
  \A^3 &=& - A_+ \sin 2 \varphi_0 \cos 2 \psi - A_\times \cos 2 \varphi_0 \sin 2 \psi \,, \no \\
  \A^4 &=& - A_+ \sin 2 \varphi_0 \sin 2 \psi + A_\times \cos 2 \varphi_0 \cos 2 \psi \,,
\end{eqnarray}
It is important to note that the $\A^\mu$ are detector independent;
for multiple detectors they are defined in a common coordinate system
(geocentric or solar system barycenter).  The detector dependent basis
signals $h_\mu^I (t)$ are defined by,
\begin{eqnarray}
  h_1^I(t) = a^I \eta(t)\cos 2\varphi(t_I)\,,\qquad h_2^I = b^I\eta(t)\cos 2\varphi(t_I)\,, \\
  h_3^I(t) = a^I \eta(t)\sin 2\varphi(t_I)\,,\qquad h_4^I = b^I\eta(t)\sin 2\varphi(t_I)\,.
\label{basis}
\end{eqnarray}
where $t_I$ is the retarded time in detector $I$. It is conventional
in the CBC literature to define $h_o(t) = \eta(t)\cos 2\varphi(t)$ and
$h_{\pi/2}= \eta(t)\sin 2\varphi(t)$.

The data in detector $I$ with signal present is:
\begin{equation}
x^I (t) = h^I (t) + n^I (t) \,,
\end{equation}
where $n^I (t)$ is the noise in detector $I$. The multi-detector data
vector is then $\x (t)$. A scalar product can be defined for each
detector $I$ on two data trains $x^I$ and $y^I$ as,
\begin{equation}
(x^I| y^I)_I = 4 \Re \left ( \int_0^\infty df~\frac{\tx^I (f) \ty^{I *} (f)}{S_n^I (f)} \right ) \,,
\end{equation}
where $S_n^I (f)$ is the one sided PSD of the noise in detector
$I$. If the noises between detectors are uncorrelated then a useful
multi-detector scalar product between vectors $\x$ and $\y$ can be
defined for the network as:
\begin{equation}
  \label{eq:innerprod-real}
  (\x| \y) = \sum_I (x^I| y^I)_I \,.
\end{equation}
We do not attach any subscript to the network scalar product - it may
be understood from the context. The multi-detector log-likelihood is
then given by:
\begin{equation}
\ln \Lambda (\A^\mu|\mathbf{x}) = (\x| \h) - \hf (\h| \h) = \A^\mu x_\mu - \hf \A^\mu \M_{\mu \nu} \A^\nu \,,
\label{LH}
\end{equation}
where
\begin{equation}
x_\mu = (\x| \h_\mu)\,,\qquad\textrm{and}\qquad \M_{\mu \nu} = (\h_\mu| \h_\nu) \,.  
\end{equation}
The log-likelihood then may be maximized with respect to the
amplitudes $\A^\mu$ to obtain the maximum likelihood or the $\Fs$
statistic:
\begin{equation}
\Fs \equiv [\ln \Lambda (\x)]_{\rm max} = \hf x_\mu \M^{\mu \nu} x_\nu \,,
\end{equation}
where $\M^{\mu \nu}$ is the inverse of $\M_{\mu \nu}$ when the inverse
exists - that is when $\M$ is non-singular. The usual coherent
multi-detector statistic is the coherent signal to noise ratio (SNR)
which is just $2 \Fs$.  That we have a maximum of $\Lambda$ rather
than an extremum requires that the eigenvalues of $M_{\mu\nu}$ be
positive definite.  We shall discuss this explicitly later.

The $\M$ matrix consists of the antenna pattern functions and the
noise variances.  For the first matrix element, consider first the CBC
case.  Here the duration of the signal can be considered short enough
that $a^I$ and $b^I$ are, to a very good approximation, constant in
time.  This approximation starts breaking down only when the duration
cannot be considered to be much smaller than a sidereal day.  Also, it
is not difficult to see that
$(h_0^I|h_0^I) \approx (h_{\pi/2}^I|h^I_{\pi/2})$, and
$(h_0^I|h_{\pi/2}^I) \approx 0$. Thus, the structure of this matrix
(in the long-wavelength limit) is:
\begin{equation}
\label{eq:Mfstat}
 \M = \left[ \begin{array}{cccc} A & C &  0 & 0 \\ 
  C &  B &  0 & 0 \\
  0 & 0 &  A & C \\ 
 0 & 0 & C & B \end{array} \right] \,,
\end{equation}  
where
$A = \mathbf{a} \cdot \mathbf{a}, B = \mathbf{b} \cdot \mathbf{b}$ and
$C = \mathbf{a} \cdot \mathbf{b}$, where $\mathbf{a}$ is the $M$
dimensional vector whose components are $a^I, ~I = 1, 2, ..., M$ and
$\mathbf{b}$ is defined similarly. The dot product is weighted by the
noise variance in each detector. For example,
${\bf a} \cdot {\bf b} := \sum_{I} \sigma_I^2 a^I b^I$ where
$\sigma_I^2 := (h_0^I|h_0^I) = (h_{\pi/2}^I|h^I_{\pi/2})$.

For CW waveforms, $\eta(t)$ is a constant, and the signals are narrow
band.  If the signal frequency is $f_0$, and in the time domain the
data duration is $T_0$ centered at the origin, then using Parseval's
inequality and that $\sin^2(2\varphi(t))$ averages to $1/2$ over many
cycles, we get for example
\begin{equation}
  (h^I_1|h^I_1)_I \approx \frac{2}{S^I_n(f_0)} \int_{-\infty}^\infty \widetilde{h}_1^{I\star}(f)\widetilde{h}_1^I(f) df \approx \frac{1}{S^I_n(f_0)}\int_{-T_0/2}^{T_0/2} (a^I(t))^2dt \,.
\end{equation}
Thus, 
\begin{equation}
  \M_{11} = \sum_I \frac{1}{S^I_n(f_0)}\int_{-T_0/2}^{T_0/2} (a^I(t))^2dt \,.
\end{equation}
It is easy to see that $\M_{\mu\nu}$ has the same form as
Eq.~(\ref{eq:Mfstat}).  The matrix elements can be written in terms of
integrals as above, which in turn can again be expressed as an inner
product as for the CBC case.

The $\B$ statistic defined in \cite{Prix:2009tq} is a Bayesian
statistic which employs a prior probability distribution on the
amplitudes. Here $\Lambda$ is a functional of the data $\x$ and
depends only on the amplitude parameters $\A^\mu$, that is,
$\Lambda (\x; \A^\mu)$ and is given by the expression in
Eq. (\ref{LH}). There is also a prior on the amplitudes which must be
supplied and we denote this by $p_{\A} (\A^\mu)$ and is a function of
the 4 amplitudes $\A^\mu$. Then the $\B$ statistic is defined as:
\begin{equation}
\B (\x) = \int \Lambda (\x; \A^\mu) p_{\A} (\A^\mu) d^4 \A  \,.
\label{Bstat}
\end{equation}
It was further shown in \cite{Prix:2009tq} that a uniform amplitude
prior on the $\A^\mu$, that is, $p_{\A} (\A^\mu) =$ const. leads one
back to the $\Fs$ statistics for a targeted search, that is, the $\B$
statistic is equivalent to the $\Fs$ statistic. As done in
\cite{Prix:2009tq}, we can assign a uniform prior on the physical
parameters $A, \cos \iota, \varphi_0, \psi$, where one then assumes that
$0 \leq A \leq A_{\rm max}$. Then a Jacobian factor enters into the
integral in Eq. (\ref{Bstat}), which is in fact the physical
prior. The Jacobian factor is:
\begin{equation}
\J = \frac{\partial (A^1, A^2, A^3, A^4)}{\partial (A, \cos \iota, \varphi_0, \psi)} = \hf A^3 (1 - \cos^2 \iota )^3 \,.
\end{equation}
and then the integrand is $\Lambda (\x; \A^\mu) / \J$, where the
integration variables are $\A^\mu$. The implicit prior in the $\Fs$
statistic is just $\J$.

\section{New complex amplitudes}
\label{sec:newamplitudes}

The standard amplitudes $\A^\mu$ defined in Eq.~(\ref{eq:amp}) are not
convenient for computing the $\B$-statistic.  For this reason a new
set of amplitudes, linear combinations of the $\A^\mu$, were defined
in \cite{Whelan:2013xka,Haris:2016jap}.  These amplitudes may be
viewed as being based on left- and right- circularly polarized signals
rather than linearly polarized signals.  We shall below define a new
set of amplitudes which are complexified versions of the amplitudes
defined in \cite{Whelan:2013xka,Haris:2016jap}.  This discussion will
cover both CW and CBC signals.

We derive these alternate complex amplitude parameters starting with
the complex strain $\mathfrak{h} = h_+ + ih_\times$, and setting
$\cos\iota = \chi$:
\begin{eqnarray}
  \mathfrak{h} &=& A \left(\frac{1+\chi^2}{2}\right)\eta(t)\cos(2\varphi_0 + 2\varphi(t)) + i A \chi\eta(t)\sin(2\varphi_0 + 2\varphi(t)) \\
  &=&  A e^{2i\varphi_0}\frac{(1+\chi)^2}{4}\mathfrak{h}_0(t) + Ae^{-2i\varphi_0}\frac{(1-\chi)^2}{4}\mathfrak{h}_0^\star(t)
\end{eqnarray}
where 
\begin{equation}
  \mathfrak{h}_0(t) = \eta(t)e^{2i\varphi(t)}\,.
\end{equation}
Combining this with the complex beam pattern function according to
Eq.(\ref{eq:hf}) yields
\begin{eqnarray}
  \label{eq:waveform-basis}
  h^I = \sum_{\mu=1}^4\B^\mu \mathfrak{h}_\mu^I
\end{eqnarray}
where 
\begin{eqnarray}
  \label{eq:newamps}
  \B_1 &=& A~e^{-2 i \varphi_0} ~\frac{(1 + \chi )^2}{4}~ e^{-2 i \psi} \,, \qquad
           \B_2 = A~e^{-2 i \varphi_0} ~\frac{(1 - \chi )^2}{4}~ e^{2 i \psi} \,, \nonumber \\
  \B_3 &=& A~e^{2 i \varphi_0} ~\frac{(1 - \chi )^2}{4} ~e^{-2 i \psi} \,, \qquad
           \B_4 = A~e^{2 i \varphi_0}~ \frac{(1 + \chi )^2}{4}~ e^{2 i \psi} \,.
\end{eqnarray}
The detector dependent complex basis functions $\mathfrak{h}_\mu^I$
are obtained by using the equations Eq. (\ref{eq:HF_basis}) and
Eq. (\ref{eq:waveform-basis}). These complex basis functions are:
\begin{equation}
  \label{eq:complexbasis}
  \mathfrak{h}_1^I = \mathfrak{h}_4^{I\star} = \hf \mathfrak{f}^I\mathfrak{h}_0^{I\star}\,,\qquad \mathfrak{h}_2^I = \mathfrak{h}_3^{I\star} = \hf \mathfrak{f}^{I\star}\mathfrak{h}_0^{I\star}\,.
\end{equation}
Since $\B_3=\B_2^\star$ and $\B_4 = \B_1^\star$, we could work just
with $\B_{2}$ and $\B_{2}$.  It shall however be convenient to
consider all the four $\B$'s when discussing the transformation from
the four real amplitudes.  As noted earlier for the beam pattern
function, these are related to the rotation matrix group elements, and
this is a reflection of the fact that, just as for the detector
response function, one should view $\varphi_0,\psi,\chi$ as
coordinates on the group of rotations.  The relations above can be
easily inverted
\begin{eqnarray}
  \chi &=& \frac{\sqrt{|\B_1|} - \sqrt{|\B_2|}}{\sqrt{|\B_1|} + \sqrt{|\B_2|}}\,,\qquad \sqrt{A} = \sqrt{|\B_1|} + \sqrt{|\B_2|}\\
  \psi &=& -\frac{1}{4}\textrm{arg}\left( \frac{\B_1}{\B_2}\right)\,,\qquad \psi = -\frac{1}{4}\textrm{arg}\left( \frac{\B_1}{\B_2^\star}\right) \,.
\end{eqnarray}
One can easily verify that the $\B_\mu$ are linear combinations of
$\A_\mu$ and vice-versa. Writing the $\B_\mu$ and $\A_\mu$ as column
vectors $\Bc$ and $\Ac$, we find that $\Ac = \S \Bc$, where $\S$ is a
$4 \times 4$ matrix given by:
\begin{equation}
\S = \hf \left[ \begin{array}{cccc} 1 & 1 &  1 & 1 \\ 
  i &  -i &  i & -i \\
  -i & -i &  i & i \\ 
 1 & -1 & -1 & 1 \end{array} \right] \,.
\end{equation}   
The determinant of $\S$ is unity and $\S$ is unitary so that
$\S^\dagger \S = I$, where $I$ is a $4 \times 4$ unit matrix;
$\S^\dagger = (\S^\intercal)^*$.  Thus we have $\Bc = \S^\dagger \Ac$.
\par

We further note that $\B_4 = \B_1^*$ and $\B_3 = \B_2^*$. This fact is
useful because when we finally write the integral for the $\B$
statistic. The volume element
$d\B_1 d\B_2 d\B_3 d\B_4 = d\B_1 d\B_1^* d\B_2 d\B_2^*$, which is then
a product of two area elements. If we assume again a uniform prior in
the physical variables $A, \chi, \varphi_0, \psi$, we will also need the
Jacobian in terms of the new amplitude variables
$\B_1, \B_2, \B_3, \B_4$.  In fact the Jacobian is unchanged because
$\S$ is unitary and we have the result:
\begin{equation}
\J = \frac{\partial (\B_1, \B_2, \B_3, \B_4)}{\partial (A, \chi, \varphi_0, \psi)} 
   = \hf A^3 (1 - \chi^2 )^3 
   \equiv 32 (\B_1 \B_2 \B_3 \B_4)^{3/4} \,.
\label{J}
\end{equation}
The Jacobian in terms of the new amplitudes is remarkably
just a product of the $\B_\mu$.

\section{Likelihood in terms of the complex amplitudes}
\label{sec:likelihood}

We begin with the definition of a suitable inner-product for two
complex signals $x(t)$ and $y(t)$ in a detector labeled by the index
$I$; $\widetilde{x}(f)$ and $\widetilde{y}(f)$ will denote their
Fourier transforms.  The usual definition of
Eq.~(\ref{eq:innerprod-real}) works only for real signals.  Let
$S^I(f)$ now be the \emph{double-sided} power-spectral density of the
detector noise.  For real-valued signals $x(t)$, for which
$\widetilde{x}^\star(f) = \widetilde{x}(-f)$ it is conventional to
only focus on positive frequencies and work instead with the
single-sided power spectral density.  However, since we need to deal
with complex functions, both positive and negative frequencies need to
be considered. The inner product of $(x|y)$ of $x$ and $y$ is then
defined as
\begin{equation}
  (x|y)_I := \int_{-\infty}^\infty \frac{\widetilde{x}^\star(f)\widetilde{y}(f)}{S^I_n(f)} df\,.
\end{equation}
Note that this is a complex inner product so that, for example, $(x|y)_I = (y|x)_I^\star$
and $(c_1y_1+c_2y_2 | x)_I = c_1^\star(y_1|x)_I + c_2^\star(y_2|x)_I$.
Here $c_{1,2}$ are complex numbers and $x,y,y_{1,2}$ are complex time
series. Note that the inner product is specific for a particular
detector through its noise spectral density.  Consider a collection of
$M$ detectors with uncorrelated noises labeled by $I=1\ldots M$.  Let
$x_I(t)$ be time series in each of the detectors which will
collectively be denoted in boldface $\mathbf{x}$.  Then, the
multi-detector inner product is
\begin{equation}
  (\mathbf{x}|\mathbf{y}) := \sum_{I=1}^M (x_I|y_I)_I = \sum_{I=1}^M \int_{-\infty}^\infty \frac{\widetilde{x}^\star_I(f)\widetilde{y}_I(f)}{S_n^I(f)} df\,.
\end{equation}
For real signals, this inner product is equivalent to
Eq.~(\ref{eq:innerprod-real}), which is why we use the same notation
for both inner products.

In terms of the $\B^\mu$, and the signal decomposition of
Eq.~(\ref{eq:waveform-basis}) it is apparent that the log-likelihood
function is still quadratic:
\begin{equation}
  \label{eq:loglambda-B}
  \ln\Lambda = \sum_{\mu=1}^4 \B^\mu (\mathbf{x}|\mathbf{\mathfrak{h}}^\mu)  - \frac{1}{2} \B^{\mu\star}\B^{\nu}N_{\mu\nu} 
             = \mathbf{B}^\dagger \mathbf{Y} - \frac{1}{2}\mathbf{B}^\dagger \mathbf{N}\mathbf{B}\,.
\end{equation}
% \jnote{To agree with Eq.~(\ref{eq:q}) later, I think that the above is
%   incorrect. Instead, we should have:
%   \begin{equation}
% \ln\Lambda = \mathbf{B}^\dagger \mathbf{Y} - \frac{1}{4}\mathbf{B}^\dagger \mathbf{N}\mathbf{B}\,.
%   \end{equation}
% Moreover, we need to define:
% \begin{equation}
%   \mathbf{B} =  \left[\begin{array}{c}
% \B_1 \\
% \B_2 \\
% \B_2^{\star} \\
% \B_1^{\star}
% \end{array}
% \right], \qquad
% \mathbf{Y} =  \left[\begin{array}{c}
% y_1 \\
% y_2^{\star} \\
% y_2 \\
% y_1^{\star}
% \end{array}
% \right]
% \end{equation}
% We may also want to consider switching to the $\B_1, \B_2, \B_1^{\star},
% \B_2^{\star}$ variables sooner, and possibly swapping the definitions of
% conjugate for $\B_2$.
% }

Here $\mathbf{Y}$ refers to the vector formed by the
$y_\mu := \sum_I(x^I|{\mathfrak{h}}^I_\mu) = (\x | \mathfrak{h}_\mu )$, and
$N_{\mu\nu}:= \sum_I (\mathfrak{h}_\mu^I|\mathfrak{h}_\nu^I)$.  We may
write out $y_\mu$ explicitly in terms of $x_\mu$:
\begin{equation}
\left[\begin{array}{c} y_1  \\ y_2 \\ y_3 \\ y_4 \end{array}\right] = \hf \left[ \begin{array}{c} x_1 - ix_2 + i x_3 + x_4 \\ x_1 + i x_2 + i x_3 - x_4 \\ x_1 - i x_2 - i x_3 - x_4 \\ x_1 + i x_2 - i x_3 + x_4 \end{array} \right]
\label{x2y}
\end{equation}  
From the above it is clear that $y_4 = y_1^*,~y_3 = y_2^*$. 

Now we come to the quadratic term in the amplitudes.  From the
definitions of the basis functions given in
Eq.~(\ref{eq:complexbasis}), it is easy to directly compute the matrix
$N_{\mu\nu}$.  For example, for the CBC case we will have,
\begin{equation}
  N_{11} = \sum_{I=1}^M (\mathfrak{h}_1^I|\mathfrak{h}_1^I)_I = \qtr \sum_{I=1}^M (\mathfrak{f}^I\mathfrak{h}_0^{I \star}|\mathfrak{f}^I \mathfrak{h}_0^{I \star})_I = \hf \sum_{I=1}^M \sigma_I^2|\mathfrak{f}^I|^2\,,
\end{equation}
where $\Vert \mathfrak{h}_0^I \Vert^2_I := 2 \sigma_I^2$ and
$|\mathfrak{f}^I|^2$ is the magnitude of the complex number
$\mathfrak{f}^I$.  Similarly, it is easy to see that,
\begin{equation}
  N_{12} = N_{21}^\star = \sum_{I=1}^M (\mathfrak{h}_1^I|\mathfrak{h}_2^I)_I = \qtr \sum_{I=1}^M (\mathfrak{f}^I\mathfrak{h}_0^\star|\mathfrak{f}^{I\star}\mathfrak{h}_0^\star)_I = \hf \sum_{I=1}^M \sigma_I^2 (\mathfrak{f}^{I\star})^2\,.
\end{equation}
Note that here we have the square of $\mathfrak{f}^{I\star}$ rather than its
norm.  It is also easy to see for example that $N_{13} = N_{14}=0$ and
in fact, $\mathbf{N}$ has the usual block diagonal form: 
\begin{equation}
\N =  \hf \left[ \begin{array}{cccc} \zeta & \kappa^\star &  0 & 0 \\ 
  \kappa &  \zeta &  0 & 0 \\
  0 & 0 &  \zeta & \kappa^\star \\ 
 0 & 0 & \kappa & \zeta \end{array} \right] \,,
\label{cov}
\end{equation}   
Obtaining this block diagonal form is a useful, and perhaps
surprising, benefit of using the complex amplitudes $\B^\mu$ rather
than the real amplitudes of \cite{Whelan:2013xka,Haris:2016jap} where
the corresponding matrix is more complicated.  The quantities $\zeta$
and $\kappa$ appearing here are:
\begin{equation}
  \zeta = \sum_{I=1}^M \sigma_I^2 |\mathfrak{f}^I|^2\,,\qquad \kappa = \sum_{I=1}^M \sigma_I^2 (\mathfrak{f}^I)^2\,.
\end{equation}
One observes that $\zeta$ is real, while $\kappa$ is complex in
general.  It is illustrative to re-express this in terms of the
$\psi$-independent beam pattern functions $a^I, b^I$ defined in
Eq.~(\ref{eq:ab}):
\begin{equation}
\label{eq:zk-ab}
  \zeta = \sum_{I=1}^M \sigma_I^2\left((a^I)^2 + (b^I)^2\right)\,,\qquad \kappa = \sum_{I=1}^M\sigma_I^2 (a^I+ib^I)^2\,.
\end{equation}
% \jnote{I believe the definitions of $\zeta$ and $\kappa$ were reversed above in Eq.~(\ref{eq:zk-ab}), so
%   I swapped them.}
For detectors with identical PSDs, the $\sigma_I$ will be identical as
well.  It is easy to see that $\zeta$ and $\kappa$ are given by:
\begin{equation}
  \zeta = | \mathbf{a} |^2 + |\mathbf{b} |^2\,,\qquad\kappa = (\mathbf{a} + i \mathbf{b})^2 = |\mathbf{a} |^2 - |\mathbf{b} |^2 + 2 i \mathbf{a} \cdot \mathbf{b} \,.
\end{equation}
Here $\mathbf{a}$ is the vector $(a^1,a^2,\cdots,a^M)$ (and similarly
for $\mathbf{b}$). The inner product is, as before, defined with the
$\sigma_I^2$ as weights.  Thus, for example,
$\mathbf{a}\cdot\mathbf{b} = \sum_I \sigma_I^2 a^I b^I$ and
$|\mathbf{a}|^2 := \mathbf{a}\cdot \mathbf{a}$.

\section{The $\B$ statistic and its evaluation in the high SNR limit}
\label{sec:evaluatingB}

We switch to the two complex variables $\B_1$ and $\B_2$ and write the
log-likelihood function as:
\begin{eqnarray}
\label{eq:q}
  \ln \Lambda &=& \B_1^* y_1 + \B_1 y_1^* + \B_2 y_2 + \B_2^* y_2^* - \hf \zeta (|\B_1|^2 + |\B_2|^2) \nonumber \\
              &-& \hf \B_1 \B_2^* \kappa - \hf \B_1^*\B_2 \kappa^*  \equiv Q (\Y; \Bc) \,. \label{eq:Q}
\end{eqnarray}
For the case $\kappa = 0$, the log likelihood can be written as sum of
two terms as was shown in \cite{Whelan:2013xka,Haris:2016jap}. In this case we can
write $\ln \Lambda = \ln \Lambda_1 + \ln \Lambda_2$, where,
\begin{equation}
  \ln \Lambda_1 = \B_1 y_1^* + \B_1^* y_1 - \hf \zeta |\B_1|^2,~~\ln \Lambda_2 = \B_2 y_2^* + \B_2^* y_2 - \hf \zeta |\B_2|^2 \,.
\end{equation}
For a uniform prior in the physical coordinates $A, \chi, \varphi_0, \psi$,
the Jacobian factor is as given in Eq. (\ref{J}) and therefore, the
$\B$ statistic is given by the integral:
\begin{equation}
\label{eq:bstat-bvars}
\B (\Y) = \frac{1}{32} \int d\B_1~ d\B_1^*~ d\B_2 ~d\B_2^* ~\left [\frac{e^{Q(\Y; \Bc)}}{({|\B_1|^2 |\B_2|^2})^{3/4}} \right] \,,
\end{equation} 
where $Q(\Y; \Bc)$ is given from Eq. (\ref{eq:Q}). The integration range
is determined from the range of $A$ - this will be better seen when we
write the same integral in polar coordinates - the limit then is on
$|\B_1|$ and $|\B_2|$.  When $\kappa = 0$, the above integral for this
prior, because this prior factorizes similarly, also can be written as
a product of two integrals, first one in $\B_1, \B_1^*$ and the second
one in $\B_2, \B_2^*$. This agrees with the result in
\cite{Whelan:2013xka}.

\subsection{The $\B$ statistic integral in polar coordinates}
\label{polar}
  
We now go over to polar coordinates in which the integral appears more
tractable. We write:
\begin{equation}
\B_1 = R_1 e^{i \theta_1}, ~ \B_2 = R_2 e^{i \theta_2}, ~y_1 = \rho_1 e^{i \phi_1}, ~y_2 = \rho_2 e^{i \phi_2},~ \kappa = k e^{4i \eta} \,.
\end{equation}
Then the log likelihood can be written as follows:
\begin{eqnarray}
\ln \Lambda &=& - \hf \zeta (R_1^2 + R_2^2) + 2 R_1 \rho_1 \cos (\theta_1 - \phi_1) + 2 R_2 \rho_2 \cos (\theta_2 - \phi_2) \no \\
&& - k R_1 R_2 \cos (\theta_1 - \theta_2 + 4\eta) \,.
\end{eqnarray}
The area element $d\B_1 d\B_1^* = - 2 i R_1 dR_1 d \theta_1$ and
similarly for $d\B_2 d\B_2^*$. The integral now acquires a minus sign
which can be corrected if we take the order of the physical variables
as $(A, \phi, \chi, \psi)$ instead of $(A, \chi, \phi, \psi)$ (Or we
can just take the modulus, if we do not care about the sign). With
this change, the $\B$ statistic is:
\begin{eqnarray}
&&\B (\rho_1, \phi_1, \rho_2, \phi_2) = \frac{1}{8} \int dR_1 dR_2 d \theta_1 d \theta_2 \times \, \no \\ 
&& \frac{e^{- \hf \zeta (R_1^2 + R_2^2) + 2 R_1 \rho_1 \cos (\theta_1 - \phi_1) +  2 R_2 \rho_2 \cos (\theta_2 - \phi_2) - k R_1 R_2 \cos (\theta_1 - \theta_2 + 4\eta)}}{\sqrt{R_1 R_2}} \,.
\label{Bpol}
\end{eqnarray}
The range over which the integral is to be carried out can be worked
out from the range allowed for the variables $(A, \chi, \phi,
\psi)$.
Clearly, since $\phi, \psi$ cover the full range, so do the angles
$\theta_1, \theta_2$ and so we have
$0 \leq \theta_1, \theta_2 < 2 \pi$. The range of $R_1, R_2$ is
decided by the range of the amplitude $A$. If $A$ is restricted to
some range $A_{\min} < A < A_{\max}$, then only $A_{\max}$ matters and
then we have $0 < R_1, R_2 < A_{\max}$. If there is no upper limit on
$A$, then the integration region is a Cartesian product of two complex
planes. Note that the integral is not singular, although the integrand
is singular at the origin, because the singularity is weak.

\subsection{The special case of $\kappa = 0$}
\label{special}

When $\kappa = 0$, the integral in Eq. (\ref{Bpol}) can be written as
a product of two integrals, one in $\rho_1, \phi_1$ and the other in
$\rho_2, \phi_2$. Thus $\B = {\rm const.} \times I_1 I_2$, where each
of the integrals are of the form:
\begin{equation}
I = \int_{0}^{\infty} dR~~ \frac{e^{- \hf \zeta R^2}}{\sqrt{R}} \int_0^{2 \pi} d \theta ~e^{2 \rho R \cos \theta} \,,
\end{equation}
where $\rho$ and $\zeta$ are constants. The integral over $\theta$ is the modified Bessel function $I_0$ and thus the integral can be written as,
\begin{equation}
I = 2 \pi \int_{0}^{\infty} dR ~~\frac{e^{- \hf \zeta R^2} I_0 (2 \rho R)}{\sqrt{R}} \,.
\end{equation}
This agrees with the expression in \cite{Whelan:2013xka} of
Eq. (6.9). As shown therein, the integral can be written in terms of a
confluent hypergeometric function.
\par

In the limit of large $\rho R$ as will be the case if the threshold is
high, the Bessel function can be replaced by its asymptotic form,
which is valid when the argument of the Bessel function is greater
than say, even $5$. In the limit $x >>1$, we have
$I_0 (x) \sim e^x /\sqrt{2 \pi x}$, we use this asymptotic form to
obtain:
\begin{eqnarray}
I &\simeq& \sqrt{\pi /\rho} \int_{R_0}^{\infty} dR~~\frac{e^{- \hf \zeta R^2 + 2 \rho R}}{R} \,, \no \\
  & = & \sqrt{\pi /\rho}~~ e^{\frac{2 \rho^2}{\zeta}}~~ \int_{R_0}^{\infty} dR~~\frac{e^{- \hf \zeta (R - 2 \rho/\zeta)^2}}{R} \,.
\end{eqnarray}
where the asymptotic form of the $I_0$ is valid for $R > R_0$. Also it
is assumed that the contribution to the integral is negligible for
$R < R_0$. Further, the $R$ in the denominator can be replaced by its
mean value in the Gaussian, namely, $2 \rho/\zeta$ and then the
Gaussian integral performed assuming
$2 \rho/\zeta - R_0 >> (\zeta)^{-1/2}$ to yield,
\begin{equation}
I \sim \pi ~ \left (\frac{\zeta}{2} \right)^{1/2} ~\rho^{-3/2}~ e^{\frac{2 \rho^2}{\zeta}} \,.
\end{equation}
Collecting all the terms and including the constant, the $\B$
statistic is given by,
\begin{equation}
\B (\rho_1, \rho_2) \simeq \frac{\pi^2}{16} ~\zeta ~[\rho_1 \rho_2]^{-3/2}~ e^{\frac{2}{\zeta}(\rho_1^2 + \rho_2^2)} \,.
\label{spl}
\end{equation} 
Recall that $\rho_1 = |y_1|$ and $\rho_2 = |y_2|$. This must agree
with the results obtained in \cite{Whelan:2013xka} if one takes the
asymptotic form of the confluent hypergeometric function as obtained
therein.

% If there is an uniform prior on $\A^\mu$, then we just get back the
% $\Fs$ statistic. Then the term $[\rho_1 \rho_2]^{-3/2}$ is absent in
% Eq. (\ref{spl}) and we are left essentially with the exponential
% term. Also since $\kappa = 0$, we have in the matrix $\Mm$ the
% entries, $A = B = |\Fp|^2 = |\Fc|^2 = \zeta/2$ and
% $C = \Fp \cdot \Fc = 0$. Thus
% $\M_{\mu \nu} = (\zeta/2) \delta_{\mu \nu}$ and its inverse
% $\M^{\mu \nu} = (2/\zeta) \delta^{\mu \nu}$. Thus we have:
% \begin{equation}
%   \Fs = \hf x_\mu \M^{\mu \nu} x_{\nu} = \hf \left(\frac{2}{\zeta} \right) x_\mu \delta^{\mu \nu} x_\nu = \hf \left(\frac{2}{\zeta} \right) (x_1^2 + x_2^2 + x_3^2 + x_4^2) \,.
% \end{equation}
% But from Eq. (\ref{x2y}) we obtain:
% \begin{equation}
%   \rho_1^2 + \rho_2^2 = |y_1|^2 + |y_2|^2 = \hf (x_1^2 + x_2^2 + x_3^2 + x_4^2) \,.
% \end{equation}
% From the above relations, we see that
% $\Fs = (2/\zeta) (\rho_1^2 + \rho_2^2)$ and therefore we have
% $\B (\x) \propto e^{\Fs (\x)}$.

\subsection{The general case of $\kappa \neq 0$}

Here the basic technique involves resorting to the principle axes
transformation which diagonalizes the matrix $\N$ given in
Eq. (\ref{cov}) and makes the quadratic form $\Bc^\dagger \N \Bc$ a
sum of squares.  We observe here that $\N$ is block diagonal
with the {\it same} block repeated. Thus we need to just focus on one
block which is just a $2 \times 2$ matrix - we call this matrix
$\N_2$:
\begin{equation}
  \N_2 = \left[ \begin{array}{cc} \zeta & \kappa^* \\ 
                  \kappa & \zeta \end{array} \right] \,,
\end{equation}
Recalling $\kappa = k e^{4i \eta}$, the eigenvalues of $\N_2$ are given
by $\zeta \pm k$. The matrix $\U$ diagonalizing $\N_2$ is unitary and
is given by:
\begin{equation}
  \U = \frac{1}{\sqrt{2}} \left[ \begin{array}{cc} e^{-2i \eta} &  -e^{-2i \eta} \\ 
                                   e^{2i \eta} & e^{2i \eta} \end{array} \right] 
                               \quad \textrm{where} \quad \tan 4\eta = \frac{2 (\mathbf{a} \cdot \mathbf{b})}{|\mathbf{a} |^2 - |\mathbf{b} |^2} \,.
\end{equation}
From the above, we have the result:
\begin{equation}
  \U^\dagger \N_2 \U = \left[ \begin{array}{cc} \zeta + k & 0 \\ 
                                0 & \zeta - k \end{array} \right] \,.
\end{equation}
The diagonalisation process leads us to another set of amplitude
variables $C_1, C_2$, where then,
\begin{equation}
  \left[\begin{array}{c} \B_1  \\ \B_2 \end{array}\right] = \U \left[ \begin{array}{c} C_1 \\ C_2 \end{array} \right] \,,
\end{equation}  
which gives,
\begin{equation}
  \B_1 = \frac{1}{\sqrt{2}}e^{-2i\eta} (C_1 - C_2) \,, \qquad \B_2 = \frac{1}{\sqrt{2}}e^{2i\eta} (C_1 + C_2) \,.  
\end{equation}
The inverse transformation is
\begin{equation}
  C_1 = \frac{1}{\sqrt{2}}\left( e^{2i\eta}\B_1 + e^{-2i\eta}\B_2 \right) \,, \qquad   C_2 = \frac{1}{\sqrt{2}}\left( -e^{2i\eta}\B_1 + e^{-2i\eta}\B_2 \right)\,.
\end{equation}
Given the expressions of Eq.~(\ref{eq:newamps}) for the $\B_\mu$, we
can view this as a (sky-dependent) shift in the polarization angle
$\psi \rightarrow \psi - \eta(\theta,\phi)$.  This is a re-derivation
of the so-called dominant polarization frame discussed in Eq.~(2.33)
of \cite{Harry:2010fr}.

Similarly, we define the complex data vectors $z_1, z_2$ by,
\begin{equation}
  \left[\begin{array}{c} z_1  \\ z_2 \end{array}\right] = \U^\dagger \left[ \begin{array}{c} y_1 \\ y_2 \end{array} \right] \,.
  \label{yz}
\end{equation}  
Therefore, $z_1 = \a y_1 + \a^* y_2$ and $z_2 = - \a y_1 + \a^* y_2$.

Now we can write the log likelihood in terms of the new amplitudes
$C_\mu$ and $z_\mu$. We then have:
\begin{equation}
  \ln \Lambda = C_1^* z_1 + C_2^* z_2 + C_1 z_1^* + C_2 z_2^* - \hf (\zeta + k) |C_1|^2 - \hf (\zeta - k) |C_2|^2 \,.
\end{equation} 
Now no cross terms in $C_1, C_2$ appear and the expression for the log
likelihood is much simpler. However, the Jacobian factor in the
denominator gets more complex because
\begin{equation}
  |B_1|^2 |B_2|^2 = \frac{1}{4} |C_1 - C_2 |^2 |C_1 + C_2 |^2 \,.
\end{equation}
The Jacobian:
\begin{equation}
\frac{\partial(B_1, B_1^*, B_2, B_2^*)}{\partial(C_1, C_1^*, C_2, C_2^*)} = 1 \,,
\end{equation}
and hence the volume elements are
$dB_1 dB_1^* dB_2 dB_2^* = dC_1 dC_1^* dC_2 dC_2^*$ since $\U$ is
unitary. We now have all the ingredients to write down the $\B$
statistic:
\begin{eqnarray}
  \label{eq:Bstat-C12}
  \B (z_\mu) &=& \frac{\sqrt{2}}{16} \int \frac{dC_1 dC_1^* dC_2 dC_2^*}{|C_1 - C_2 |^{3/2} |C_1 + C_2 |^{3/2}} \, \no \\
             &\times& e^{C_1^* z_1 + C_2^* z_2 + C_1 z_1^* + C_2 z_2^* - \hf (\zeta + k) |C_1|^2 - \hf (\zeta - k) |C_2|^2} \,.
\end{eqnarray}
We first consider the case when the signal-to-noise-ratio (SNR)
is sufficiently high so that we can focus on the peak of the terms in
the exponent. Collecting terms in $C_1$ and $C_2$, writing
$\ln\Lambda = \ln\Lambda_+ + \ln\Lambda_-$, and completing squares in
$\ln\Lambda_{\pm}$:
\begin{equation}
  \ln\Lambda_\pm = \frac{2|z|^2}{\zeta \pm k} - \frac{1}{2} (\zeta\pm k)\left| C - \frac{2z}{\zeta\pm k}\right|^2 \,.
\end{equation}
Here it is assumed that in $C$ and $z$ are respectively $C_1$ and
$z_1$ in $\ln\Lambda_+$, and $C_2$ and $z_2$ in $\ln\Lambda_2$.  Under
the simplest approximation, in the denominator of the integrand in
Eq.~(\ref{eq:Bstat-C12}), we set $C_{1,2}$ to the values at the peak
of the Gaussian 
\begin{equation}
  C_1 = \frac{2z_1}{\zeta+k}\,,\quad C_2 = \frac{2z_2}{\zeta-k}\,.
\end{equation}
Performing the two of Gaussian integrals yields
(taking the integrals to be over the whole complex plane)
\begin{equation}
  \label{eq:gnrl}
  \B = \frac{4\pi^2}{ (\zeta^2-k^2)}\left|\frac{4z_1^2}{(\zeta+k)^2} - \frac{4z_2^2}{(\zeta-k)^2}\right|^{-3/2} \exp\left[\frac{2|z_1|^2}{(\zeta+k)} + \frac{2|z_2|^2}{(\zeta-k)} \right]
\end{equation}

We first check that this expression reduces to the special case of
$\kappa = 0$ derived in section \ref{special}, Eq. (\ref{spl}).  In
this case, we put $k = 0$ in Eq. (\ref{eq:gnrl}). We first note that
$z_1, z_2$ are related to $y_1, y_2$ by a unitary transformation $\U$
given in Eq. (\ref{yz}). Taking the Hermitian conjugate of this
equation and multiplying this with the equation itself, it follows
that $|z_1|^2 + |z_2|^2 = |y_1|^2 + |y_2|^2$. This takes care of the
exponential term. For the denominator, we see that
$|z_1 - z_2| = \sqrt{2} |y_1|$ and $|z_1 + z_2| = \sqrt{2}
|y_2|$. Again collecting all terms we finally deduce that:
\begin{equation}
  \B = \frac{\pi^2}{16} ~\zeta ~ \frac{e^{\frac{2}{\zeta} \left (|y_1|^2 + |y_2|^2 \right)}}{(|y_1| |y_2|)^{3/2}} \,,
\end{equation}
which is in exact agreement with Eq. (\ref{spl}).

Also from the expression of $\kappa$, we find
$k^2 = |\kappa|^2 = |\mathbf{a}|^4 + |\mathbf{b}|^4 + 2 |\mathbf{a}|^2
|\mathbf{b}|^2 \cos 2 \theta \leq (|\mathbf{a}|^2 + |\mathbf{b}|^2)^2
\equiv \zeta^2$,
where we have written
$\mathbf{a} \cdot \mathbf{b} = |\mathbf{a}||\mathbf{b}| \cos \theta$.
This means that $k \leq \zeta$.  The $\B$ statistic is singular when
$\zeta=k$.  In fact, setting $\epsilon = \zeta-k$ with $\epsilon$
small, we see that
\begin{equation}
  \B \sim \epsilon^2 e^{2|z_2|^2/\epsilon} \,.
\end{equation}
This is singular when $\epsilon \rightarrow 0$.

\subsection{Singularities of the $\B$ statistic}
\label{subsec:singularities}

The singularities of the $\B$ statistic given by Eq. (\ref{eq:gnrl})
occur when $\zeta = k$.  We can see from Eq.~(\ref{eq:gnrl}) that it
will \emph{also} be singular when:
\begin{equation}
  \frac{z_1}{\zeta+\kappa} = \pm \frac{z_2}{\zeta-\kappa}.
\end{equation}
These correspond to the two cases where the maximum-likelihood signal
is purely circularly polarized, either left or right. Whereas the singularities
that occur when $\zeta = \pm k$ are characteristic of the detector network
alone, the singularities above depend also on the data, and the extent to which
it prefers pure circular polarization.

Let us now consider the first kind of singularity.  To avoid clutter,
we drop the factors of $\sigma_I$ which account for the detector
sensitivities (or alternatively, scale the beam pattern functions
appropriately -- this will not affect the present discussion). Given
that $\zeta \geq k$ and $\zeta,k \geq 0$, we see that $\zeta+k = 0$ if
and only if $\zeta=0$.  Consider then the quantity $\zeta^2-k^2$.  If
this quantity vanishes then either $\zeta=0$ or $\zeta=k$. Thus, away
from sky-positions where $\zeta=0$, which is equivalent to both $a$
and $b$ vanishing for each detector, we see that the set of points
where $\zeta^2-k^2$ vanishes contains all such singular points. A
straightforward calculation yields
\begin{eqnarray}
  \zeta^2-k^2 &=& \left(\sum_{I}|\mathfrak{f}_I|^2 \right)^2 - \left|\sum_I\mathfrak{f}_I^2\right|^2 
                  = \sum_{I,J}\left(\mathfrak{f}_I\mathfrak{f}_I^\star\mathfrak{f}_J\mathfrak{f}_J^\star - \mathfrak{f}_I^2\mathfrak{f}_J^{\star 2}\right) \nonumber \\
              &=& -\sum_{I>J} \left(\mathfrak{f}_I\mathfrak{f}_J^\star - \mathfrak{f}_I^\star\mathfrak{f}_J\right)^2\,. 
\end{eqnarray}
The terms with $I=J$ vanish and we note also that the quantity
$\mathfrak{f}_I\mathfrak{f}_J^\star -
\mathfrak{f}_I^\star\mathfrak{f}_J$
is pure imaginary.  Thus its square is negative and we get
\begin{equation}
  \zeta^2-k^2 = \sum_{I>J} \left|\mathfrak{f}_I\mathfrak{f}_J^\star - \mathfrak{f}_I^\star\mathfrak{f}_J\right|^2\,.
\end{equation}
We have thus written $\zeta^2-k^2$ as a sum of positive terms and as
a sum over detector pairs.  If $\zeta^2-k^2=0$, then each term in the
above sum must vanish. Consider now a single term from this expression: 
\begin{equation}
  \left| \mathfrak{f}_I\mathfrak{f}_J^\star - \mathfrak{f}_I^\star\mathfrak{f}_J\right| = |\mathfrak{f}_I\mathfrak{f}_J|\times \left|\frac{\mathfrak{f}_J^\star}{\mathfrak{f}_J} - \frac{\mathfrak{f}_I^\star}{\mathfrak{f}_I}\right|\,.
\end{equation}
We see immediately that apart from points where $\mathfrak{f}_I$ and
$\mathfrak{f}_J$ vanish, the degeneracy occurs when
$\frac{\mathfrak{f}_J^\star}{\mathfrak{f}_J} =
\frac{\mathfrak{f}_I^\star}{\mathfrak{f}_I}$.
This is just the condition that the arguments of $\mathfrak{f}_I$ and
$\mathfrak{f}_J$ are equal (or differ by $\pi$).  The condition for
$\zeta^2-k^2$ to vanish is then that the arguments for \emph{all}
$\mathfrak{f}_I$ should be equal (modulo $\pi$).  For two detectors,
we will thus generically get curves on the celestial sphere. For three
detectors, we will get a finite number of degenerate points (the
intersection of the curves), and no degeneracy generically for 4 or
more detectors.

The second kind of singularity can also be discussed in terms of the
quantity $\zeta^2-k^2$ but is more complicated due to the presence of
the data.  We shall leave further discussion of the geometric nature
of the singularity to future work and turn now to the properties of
the $\B$ statistic at the singularity.

\subsection{Applying the Laplace approximation to the $\B$ statistic}
\label{sec:laplace}

We can formalize the approximations to the integral that we have been making
through use of \emph{Laplace's approximation}.  As the name suggests, the
technique is an old one.  In its most familiar form, it states that if a
function $f$ takes on a unique, non-degenerate minimum at the origin within some
domain $R$ of $\mathbb{R}^n$ that properly contains the origin, and that if the
value of that minimum is zero, then:
\begin{equation}
  \label{eq:laplace}
  \int_R \, e^{-\lambda f}\,g(x)\,d^n \mathbf{x} \,\sim\,
  \left(\frac{2\pi}{\lambda}\right)^{n/2}\,\frac{g(\mathbf{0})}{\sqrt{\det{Hf(\mathbf{0})}}}
\end{equation}
for sufficiently large $\lambda$. Here $Hf(\mathbf{0})$ denotes the Hessian of $f$,
evaluated at the origin. This approximation, and the related
stationary-phase approximation, are well known in applied mathematics and we
have used them already in this paper. They encapsulate the idea that for a
sharply peaked integrand, the value of the integral should be chiefly determined
by its behavior at its peak.

Much less well known is that the approximation of Eq.~(\ref{eq:laplace}) is just
the first term in a full asymptotic expansion.  Moreover, the requirements that
the minimum be non-degenerate, and the evident requirement that $g$ be
well-defined at the origin, may both be relaxed.  All that is really required is
that both $f$ and $g$ themselves possess asymptotic expansions, as $|\mathbf{x}|
\rightarrow \mathbf{0}$, and that any singularities of $g$ should be integrable. There is
extensive discussion of these topics in the book by
Wong~\cite{wong2001asymptotic}, but we shall refer primarily to the
paper~\cite{Kirwin:2010} by Kirwin, which collects in one place the specific
generalizations of Laplace's method that we shall need.

We will focus only on the leading order term in the asymptotic expansion; our
interest in generalizing is to relax the assumption that the prior (that part of
the integrand not in the exponential) be finite: we will instead allow it an
integrable singularity. As we shall see, this is necessary to apply Laplace's
approximation to the case where the gravitational wave strain data is peaked at
a pure circular polarization state.

So, we still assume that the function $f$ in Eq.~(\ref{eq:laplace}) has a unique
minimum value at the origin, and that the value of that minimum is zero.  We do
\emph{not} assume that the minimum is non-degenerate. If $\mathbf{x} =
(x^1,\ldots,x^n)$ are Cartesian coordinates on $\mathbb{R}^n$, then introduce generalized
spherical coordinates by defining  $r = \sqrt{(x^1)^2+\cdots(x^n)^2}$, and
denoting by $S^{n-1}$ the unit sphere $r = 1$; also define $\Omega =
\mathbf{x}/r$. The hypotheses that are required to apply the results
of~\cite{Kirwin:2010} are: 
\begin{enumerate}
\item The function $f$ possesses $N+1$ continuous functions $f_j$ on $\Omega$,
  with $f_0(\Omega) > 0$, such that for some real number $\nu > 0$:
  \begin{equation}
   f(r,\Omega) = r^{\nu} \sum_{j=0}^{N} f_j(\Omega) r^j + o(r^{N+\nu}) \qquad\mbox{ as
     $r\rightarrow 0$, and }
  \end{equation}
\item The function $g$ possesses $N+1$ continuous functions $g_j$ on $\Omega$,
  such that for some real number $\mu > 0$:
  \begin{equation}
   g(r,\Omega) = r^{\mu-n} \sum_{j=0}^{N} g_j(\Omega) r^j + o(r^{N+\mu-n}) \qquad\mbox{ as
     $r\rightarrow 0$, and}
  \end{equation}
\item There exists some $\lambda_0 > 0$ such that the integral
  \begin{equation}
   \int_R e^{-\lambda_0 f}\,g\,d^nx
  \end{equation}
converges.
\end{enumerate}
When these conditions hold, the leading term of the asymptotic expansion is
given by:
\begin{equation}
  \label{eq:extended-laplace}
   \int_R e^{-\lambda f}\,g\,d^nx
   \,\sim\, \lambda^{-\mu/\nu}\frac{1}{\nu}\Gamma\left(\frac{\mu}{\nu}\right)\int_{S^{n-1}}
   \frac{g_0(\Omega)}{\left[f_0(\Omega)\right]^{\mu/\nu}} \,d\Omega
\end{equation}
Though it is not obvious from Eq.~(\ref{eq:extended-laplace}), in the familiar
case where the minimum of $f$ is non-generate (implying the determinant of the
Hessian of $f$ at its minimum is nonzero) and when $\mu = n$, so that $g$ has no
singularity at the minimum of $f$, then Eq.~(\ref{eq:extended-laplace}) reduces
to Eq.~(\ref{eq:laplace}).

To apply either Eq.~(\ref{eq:laplace}) or Eq.~(\ref{eq:extended-laplace}), we
may use either integral form of the $\B$ statistic, whether Eq.~(\ref{eq:bstat-bvars}) or
Eq.~(\ref{eq:Bstat-C12}). In either case, we must rewrite the expression so that
the minimum of our log-likelihood occurs at the origin, and the value of that
minimum must be zero. This is accomplished in each case by completing the
square. 

If we start with Eq.~(\ref{eq:bstat-bvars}), define:
\begin{equation}
  \label{eq:hatb}
  \hat{\Bc} = \N^{-1}\Y, \qquad \Delta\Bc = \Bc-\hat{\Bc}
\end{equation}
Then it is easy to show that:
\begin{equation}
  \label{eq:qcenter}
  Q(\Y;\Bc) = -\frac{1}{2}\Delta\Bc^{\dagger} \N \Delta\Bc + \frac{1}{2}\hat{\Bc}^{\dagger}
  \N \hat{\Bc}
\end{equation}
Using this, we get:
\begin{equation}
  \label{eq:bcentered}
  \B(\Y) = \frac{e^{\frac{1}{2}\hat{\Bc}^{\dagger}\N\hat{\Bc}}}{32}\, \int
    \frac{d(\Delta\B_1)\,d(\Delta\B_1^{\star})\,d(\Delta\B_2)\,d(\Delta\B_2^{\star})\,
      e^{-\frac{1}{2}\Delta\Bc^{\dagger}\N\Delta\Bc}}{\left[(\Delta\B_1+\hat{\B}_1)(\Delta\B_1^{\star}+\hat{\B}_1^{\star})
        (\Delta\B_2+\hat{\B}_2)(\Delta\B_2^{\star}+\hat{\B}_2^{\star})\right]^{3/4}}
\end{equation}
We now have an exponential in the integrand which has its minimum value at the
origin, and a minimum value of zero. It is also helpful to explicitly include a
parameter that will be large when, as we consider, there is a threshold for
triggers. Define the following:
\begin{eqnarray*}
\lambda &= \sqrt{|\hat\B_1|^2+|\hat\B_2|^2}, \\
\mathbf{U} &= \Delta\Bc/\lambda, \\
\hat{\mathbf{U}} &= \hat{\Bc}/\lambda .
\end{eqnarray*}
Using this, we can write:
\begin{equation}
  \label{eq:blaplace}
  \B(\Y) = \lambda\frac{e^{\frac{1}{2}\hat{\Bc}^{\dagger}\N\hat{\Bc}}}{32}\, \int
    \frac{d(U_1)\,d(U_1^{\star})\,d(U_2)\,d(U_2^{\star})\,
      e^{-\frac{\lambda^2}{2}\mathbf{U}^{\dagger}\N\mathbf{U}}}{\left[(U_1+\hat{U}_1)
        (U_1^{\star}+\hat{U}_1^{\star})
        (U_2+\hat{U}_2)(U_2^{\star}+\hat{U}_2^{\star})\right]^{3/4}}
\end{equation}
Note that $|\hat{U}_1|^2 + |\hat{U}_2|^2 = 1$. 

It is now straightforward to consider the cases of Laplace's transformation that
we may apply to this expression. If neither of $|\hat{U}_1|$ or $|\hat{U}_2|$
are zero, then the function $g$ in our formulation of Laplace's approximation is
not singular at the origin, and we may directly apply Eq.~(\ref{eq:laplace}) to
get: 
\begin{equation}
  \label{eq:bnonsing}
  \B(\Y) \sim \frac{\pi^2}{2(\zeta^2-k^2)}\frac{e^{\frac{1}{2}\hat{\Bc}^{\dagger}\N\hat{\Bc}}}{\left(|\hat{\B}_1||\hat{\B}_2|\right)^{3/2}}.
\end{equation}
and where we remind the reader that the functional dependence of the right-hand
side on $\Y$ is through $\hat{\Bc} = \N^{-1}\Y$.

However, if the peak value of one of $|\hat{U}_1|$ or $|\hat{U}_2|$ does vanish,
then we must use Eq.~(\ref{eq:extended-laplace}); this happens if the signal is
purely circularly polarized. For concreteness, suppose that $\hat{U}_1 = 0$;
note that this then implies $|\hat{U}_2| = 1$. In that case, as $r\rightarrow
0$, the factors in $g$ that contain $\hat{U}_1$ diverge, though the singularity
is integrable. We have:
\begin{eqnarray}
  \label{eq:f0}
  f(r, \Omega) &= \mathbf{U}^{\dagger} \N \mathbf{U} \\ \nonumber
   &=  r^2\[\zeta + 2k\sin\beta\,\cos\beta\cos(\alpha_1-\alpha_2+4\eta)\right]
\end{eqnarray}
In this equation, we have introduced Hopf coordinates on $S^3$:
\begin{equation}
  \label{eq:hopf}
  U_1 = r\cos\beta\,e^{i\alpha_1} \,,\quad U_2 = r\sin\beta\,e^{i\alpha_2}\,, 
 \quad d\Omega = \sin\beta \cos\beta\,d\beta\,d\alpha_1 d\alpha_2.
\end{equation}

In these same coordinates, we have:
\begin{equation}
  \label{eq:g}
  g(r,\Omega) \rightarrow
  \left[(r\cos\beta\,e^{i\alpha_1})(r\cos\beta\,e^{-i\alpha_1})\right]^{-3/4}
  = r^{-3/2}(\cos\beta)^{-3/2}
\end{equation}
From this we conclude that the exponent $\mu$ is $5/2$; it is also immediate
that the exponent $\nu$ is two, since our function $f$ is a quadratic
form. Thus, if we put together all of the elements of
Eq.~(\ref{eq:extended-laplace}), we obtain in this case:
\begin{equation}
  \label{eq:bsing}
  \B(\Y) \sim \frac{2^{1/4}}{32\zeta^{5/4}|\hat{\B}_2|^{3/2}}\Gamma\left(\frac{5}{4}\right)
  e^{\frac{1}{2}\hat{\Bc}^{\dagger}\N\hat{\Bc}} I\left(\frac{k}{\zeta}\right)
\end{equation}
where we have defined the function $I(x)$, for $0\leq x < 1$, by the
three-dimensional integral:
\begin{eqnarray}
  \label{eq:iint}
  I(x) &=  \int_0^{\pi/2} d\beta \int_0^{2\pi} d\alpha_1 \int_0^{2\pi} d\alpha_2
  \\ \nonumber
& \qquad \times \frac{\sin\beta\cos\beta}{\left[\cos\beta\right]^{3/2} \left[1+2x\sin\beta\cos\beta\cos{(\alpha_1-\alpha_2+4\eta)}\right]^{5/4}}
\end{eqnarray}
Though this integral may appear intractable, in fact it may be performed
exactly, as outlined in~\ref{sec:eval-integr-ix}. The result is:
\begin{equation}
  \label{eq:ix-eval}
  I(x) = 8\pi^2 (1-x^2)^{-1/4}
\end{equation}
and therefore we obtain:
\begin{equation}
  \label{eq:bsing-simp}
  \B(\Y) \sim \frac{2^{1/4}\Gamma\left(\frac{1}{4}\right)\pi^2}{16\zeta^{3/4}(\zeta^2-k^2)^{1/4}}\frac{
  e^{\frac{1}{2}\hat{\Bc}^{\dagger}\N\hat{\Bc}}}{|\hat{\B}_2|^{3/2}}
\end{equation}

If we compare this equation to Eq.~(\ref{eq:bnonsing}), we see that the
exponential factor in Eq.~(\ref{eq:bsing-simp}) is identical to that in
Eq.~(\ref{eq:bsing-simp}), as is the dependence on $\hat{\B}_2$. The only
functional difference in the dependence on the data is that the singularity that 
Eq.~(\ref{eq:bnonsing}) would predict as $\hat{\B}_1\rightarrow 0$ is absent
from Eq.~(\ref{eq:bsing-simp}). 

It would seem desirable to have an expression that smoothly interpolates between
Eqs.~(\ref{eq:bnonsing}) and (\ref{eq:bsing-simp}). This is a considerably more
complex problem. The essential difficulty is that we are attempting an
asymptotic approximation not of a single integral, but rather a family of
integrals, depending on parameters (in our case, the $\Y$ and through those the
$\hat{\Bc}$). To smoothly interpolate, we would need an asymptotic approximation
that was \emph{uniform} in these parameters.  This topic is discussed in depth
in chapter seven of Wong's book~\cite{wong2001asymptotic} and references cited
therein, as well as a different approach in~\cite{Connor1973}, but we have found
no simple application of those techniques to our problem, and defer an
investigation of that for future work.

\section{Discussion}
\label{sec:discussion}

In this paper, we have studied the analytic marginalization of the
likelihood function over the nuisance parameters
$(A,\varphi_0,\iota,\psi)$ for modeled gravitational wave searched.
The results are applicable for both transient binary coalescence
signals, and the long duration signals emitted by rapidly rotating
neutron stars.  The main theme of this paper is that the matrix
elements of the rotation group $T^\ell_{mn}$ are useful amplitude
parameters.  Not only are they natural from a geometric viewpoint, but
they also simplify the calculation of the $\B$-statistic.  The same
interpretation is useful also in the calculation of the beam pattern
function for a gravitational wave detector.  Thus, it is natural to
view the coordinates $\varphi_0,\psi,\iota$ as coordinates on the
group of rotations or equivalently, on $S^3$.  Using these
coordinates, we have derived useful approximations to the
$\B$-statistic in the case when the SNR is high.  We have investigated
the singularities of the detector network and we have obtained a
simple expression for $\B$ in the singular case.

We have restricted ourselves to non-precessing systems and in fact we
have also not considered higher modes in the waveform model.  In both
these case, we will need to go beyond the dominant mode of the
gravitational wave signal.  It is clear that these higher modes can be
included in our formalism quite naturally. It also seems plausible
that extensions of the methods used here might also work in those
cases.  It is an accident that for non-precessing systems, the number
of amplitude parameters is the same as the number of physical
parameters.  Including precession or higher modes will result in a
larger number of amplitude parameters than the number of physical
parameters.  The Bayesian framework naturally deals with this mismatch
in the number of parameters.  Even restricting ourselves to non-precessing
systems and ignoring the higher modes, the results of this paper might
be useful in parameter estimation methods and in suggesting
modifications to the detection statistic for the searches as well. 

\ack

We are grateful to Archana Pai for valuable comments.  JLW
acknowledges support from NSF award PHY-1506254.

\section*{References}

\bibliography{bstat_cqg}

\appendix

\section{The matrix elements of the $\ell=2$ representation of the rotation group}
\label{sec:appendix}

A rotation $g$ is represented by matrices $T^\ell_{mn}[g]$ where
$\ell$ is the weight of the representation and $m,n$ are the indices
for the matrix elements with $-\ell \leq m,n \leq \ell$.  If we use
the Euler angles $(\alpha,\beta,\gamma)$ to parametrize a rotation
then $T^\ell_{mn}$ are functions of $(\alpha,\beta,\gamma)$.
Following \cite{GMS} we shall use the 'zxz' convention for the Euler
angles.  Thus we shall go from a frame $(x,y,z)$ to
$(x^\prime,y^\prime,z^\prime)$ starting with a rotation around the $z$
axis by an angle $\alpha$, a rotation around the $x$-axis by an angle
$\beta$, and finally a rotation around the $z$ axis by an angle
$\gamma$. The $T^{\ell}_{mn}$ are also called the Wigner D-matrices,
and the spin weighted spherical harmonics
${}_sY_{\ell m}(\theta,\phi)$ are proportional to
$T^\ell_{-s,m}(\theta,\phi,0)$, i.e. with the third Euler angle set to
zero \cite{Goldberg:1966uu}.

It can be shown that
\begin{equation}
  T^\ell_{mn}(\alpha,\beta,\gamma) = e^{-im\alpha}P^\ell_{mn}(\cos\beta)e^{-in\gamma}\,,
\end{equation}
where $P^\ell_{mn}(\cos\beta)$ are the Jacobi polynomials:
\begin{eqnarray}
  P^\ell_{mn}(x) &=& \frac{(-1)^{\ell-m}i^{n-m}}{2^\ell(\ell-m)!}\sqrt{\frac{(\ell-m)!(\ell+n)!}{(\ell+m)!(\ell-n)!}}   \nonumber \\
  &\times& (1-x)^{-\frac{n-m}{2}}(1+x)^{-\frac{n+m}{2}}\frac{d^{\ell-n}}{dx^{\ell-n}}[(1-x)^{\ell-m}(1+x)^{\ell+m}]\,.
\end{eqnarray}
It will be useful to list explicitly the $\ell=2$ matrix elements for $m=-2$
\begin{eqnarray}
  P^2_{-2,-2}(\cos\beta) &=&  \frac{1}{4}(1+\cos\beta)^2 \,,\\ 
  P^2_{-2,-1}(\cos\beta) &=& \frac{i}{2}\sin\beta(1+\cos\beta) \,,\\
  P^2_{-2,0}(\cos\beta) &=& -\frac{1}{2}\sqrt{\frac{3}{2}}(1-\cos^2\beta) \,,\\ 
  P^2_{-2,1}(\cos\beta) &=& \frac{i}{2}\sin\beta(\cos\beta-1) \,,\\ 
  P^2_{-2,2}(\cos\beta) &=& \frac{1}{4}(1-\cos\beta)^2  
\end{eqnarray}
Note that
\begin{equation}
  P^{\ell}_{mn}(x) = P^{\ell\,\star}_{nm}(x)\,.
\end{equation}

\section{Evaluating the integral $I(x)$}
\label{sec:eval-integr-ix}

Here we outline the steps in the closed-form evaluation of the function $I(x)$
defined in Eq.~(\ref{eq:iint}), which we repeat here for convenience:
\begin{eqnarray}
  \label{eq:iintb}
  I(x) &=  \int_0^{\pi/2} d\beta \int_0^{2\pi} d\alpha_1 \int_0^{2\pi} d\alpha_2
  \\ \nonumber
& \qquad \times \frac{\sin\beta\cos\beta}{\left[\cos\beta\right]^{3/2} \left[1+2x\sin\beta\cos\beta\cos{(\alpha_1-\alpha_2+4\eta)}\right]^{5/4}}
\end{eqnarray}
To begin our evaluation of this integral, first change variables from $\alpha_1$
and $\alpha_2$ to new variables $\gamma_{+}$ and $\gamma_{-}$ defined by:
\begin{equation}
  \label{eq:gammas}
  \gamma_{+} = \alpha_1 + \alpha_2, \qquad \gamma_{-} = \alpha_1 - \alpha_2
\end{equation}
It is easy to check that $d\alpha_1 d\alpha_2 = \frac{1}{2}d\gamma_{+}
d\gamma_{-}$; we must also adjust the limits of integration.  Our goal is to
perform the integral over $\gamma_{+}$ trivially, so we write:
\begin{eqnarray}
  I(x) &=  \frac{1}{2} \int_0^{\pi/2} d\beta \int_{-2\pi}^{2\pi} d\gamma_{-} \int_{-2\pi+|\gamma_{-}|}^{2\pi-|\gamma_{-}|} d\gamma_{+}
  \\ \nonumber
& \qquad \times \frac{\sin\beta\cos\beta}{\left[\cos\beta\right]^{3/2}
  \left[1+2x\sin\beta\cos\beta\cos{(\gamma_{-}+4\eta)}\right]^{5/4}} \\   \label{eq:gamma-int-1}
  &= \frac{1}{2} \int_0^{\pi/2} d\beta \int_{-2\pi}^{2\pi} d\gamma_{-}
    (4\pi-2|\gamma_{-}|) \\ \nonumber
& \qquad \times \frac{\sin\beta\cos\beta}{\left[\cos\beta\right]^{3/2}
  \left[1+2x\sin\beta\cos\beta\cos{(\gamma_{-})}\right]^{5/4}}   \label{eq:gamma-int-2}
\end{eqnarray} 
In going from Eq.~(\ref{eq:gamma-int-1}) to Eq.~(\ref{eq:gamma-int-2}), we have
performed the integral over $\gamma_{+}$. It is subtler to see that we may drop
the shift by $4\eta$, but it is possible to check that the potential extra term
in the $|\gamma_{-}|$ in fact vanishes when integrated from $-2\pi$ to $2\pi$.

Next, we use the identity:
\begin{equation}
  \label{eq:pow-taylor}
  (1-z)^{-a} = \sum_{k=0}^{\infty} \frac{(a)_k}{k!} z^k, \qquad |z| < 1
\end{equation}
where $(a)_k = \Gamma(a+k)/\Gamma(a)$ is the Pochhammer symbol, to expand the
binomial fractional power in the integrand; this expansion will be valid for $0
\leq x < 1$ as we require. This gives:
\begin{eqnarray}
  \label{eq:expand-pow}
  I(x) &= \frac{1}{2} \sum_{k=0}^{\infty} \frac{(\frac{5}{4})_k}{k!} (-2x)^k
         \int_{-2\pi}^{2\pi} \left(2\pi-|\gamma_{-}|\right) (\cos{\gamma_{-}})^k
         d\gamma_{-} \\ \nonumber
       & \qquad \times \int_{0}^{\pi/2} \frac{
         (\sin\beta)^{k+1}(\cos\beta)^{k+1} }{[\cos\beta]^{3/2}} d\beta.
\end{eqnarray}
The integral over $\gamma_{-}$ can now be performed by elementary methods, and
vanishes when $k$ is odd.  When $k = 2m$ is even, it is:
\begin{equation}
  \label{eq:int-gamma-minus}
\int_{-2\pi}^{2\pi} \left(2\pi-|\gamma_{-}|\right) (\cos{\gamma_{-}})^{2m}
d\gamma_{-} = \frac{4\pi^2}{2^{2m}} {2m\choose{m}}
\end{equation}
and inserting this gives:
\begin{equation}
  \label{eq:ix-oneint}
  I(x) = 4\pi^2 \sum_{m=0}^{\infty} \frac{(\frac{5}{4})_{2m}}{(2m)!}
         {2m\choose{m}} x^{2m} \int_0^{\pi/2}
         (\sin\beta)^{2m+1}(\cos\beta)^{2m-\frac{1}{2}} d\beta
\end{equation}

Now we define $u = \cos\beta$ to re-express the remaining integral; if we then
further define $u =v^2$ and expand the binomial in the integrand, we get:
\begin{equation}
  \label{eq:inner-int}
  \int_0^{\pi/2} (\sin\beta)^{2m+1}(\cos\beta)^{2m-\frac{1}{2}} d\beta 
   = 2\sum_{l=0}^{m} {m\choose{l}} \frac{(-1)^l}{4(l+m)+1}.
\end{equation}
It is perhaps not obvious that this sum can be analytically performed, but in
fact it is tractable to computer algebra systems (we used
\texttt{SymPy}~\cite{sympy}) to get: 
\begin{equation}
  \label{eq:inner-analyt}
  \int_0^{\pi/2} (\sin\beta)^{2m+1}(\cos\beta)^{2m-\frac{1}{2}} d\beta 
   = \frac{2\Gamma(m+1)\Gamma(m+\frac{5}{4})}{(4m+1)\Gamma(2m+\frac{5}{4})}.
\end{equation}
If we insert this into our expression Eq.~(\ref{eq:ix-oneint}) for $I(x)$, we
have now only a single summation and no remaining integrals:
\begin{equation}
  \label{eq:one-sum}
    I(x) = 8\pi^2 \sum_{m=0}^{\infty} \frac{(\frac{5}{4})_{2m}}{(2m)!}
         {2m\choose{m}}
         \frac{\Gamma(m+1)\Gamma(m+\frac{5}{4})}{(4m+1)\Gamma(2m+\frac{5}{4})}x^{2m}.
\end{equation}
Finally, we use the identities $n! = \Gamma(n+1)$, $\Gamma(z+1) = z\Gamma(z)$,
and the expression for the binomial coefficient to rewrite this as:
\begin{equation}
  \label{eq:one-sum-simple}
    I(x) = 8\pi^2 \sum_{m=0}^{\infty} \frac{(\frac{1}{4})_{m}}{(m)!} (x^2)^m
\end{equation}
which is recognized as the Taylor expansion of $8\pi^2\,(1-x^2)^{-1/4}$, as
claimed. Given the remarkable simplicity of this result, it would be interesting
to find a simpler derivation, but so far we have not.

\end{document}